\documentclass[prl,twocolumn,showpacs]{revtex4}

\usepackage{graphicx}
\usepackage{amsmath}
\usepackage{amsfonts}
\usepackage{amssymb}
\usepackage{epsf}
\usepackage{hyperref}

\begin{document}
\title{Transport of Atom Packets in a Train of Ioffe-Pritchard Traps}
\author{T. Lahaye, G. Reinaudi, Z. Wang, A. Couvert and D. Gu\'ery-Odelin}
\affiliation{Laboratoire Kastler Brossel$^{*}$, 24 rue Lhomond,
F-75231 Paris Cedex 05, France}
 \date{\today}

\begin{abstract}
We demonstrate transport and evaporative cooling of several atomic
clouds in a chain of magnetic Ioffe-Pritchard traps moving at a
low speed ($<1$~m/s). The trapping scheme relies on the use of a
magnetic guide for transverse confinement and of magnets fixed on
a conveyor belt for longitudinal trapping. This experiment
introduces a new approach for parallelizing the production of
Bose-Einstein condensates as well as for the realization of a
continuous atom laser.
\end{abstract}


\maketitle

The combination of laser cooling and evaporative cooling has led
in the last decade to a revolution in atomic physics, with the
achievement of Bose-Einstein Condensation (BEC) in alkali
vapors~\cite{bec}. This breakthrough was followed by many
spectacular experiments, among which the realization of coherent
atom sources~\cite{atomlasers} that are the equivalent of a laser
for matter waves. However the available mean flux of
quantum-degenerate atoms has remained limited to values between
$10^4$ and $10^6$ per second. Larger fluxes would be highly
beneficial for applications such as lithography or interferometry.

A natural way to increase the production rate of Bose-Einstein
condensates is to increase the duty cycle of laser cooling in the
magneto-optical trap (MOT), which is extremely efficient in terms
of cooling rate. For that purpose, one needs to operate the
various steps of evaporative cooling \emph{at the same time}, at a
location differing from the one where laser cooling takes place. A
first possibility, proposed in~\cite{Mandonnet00} and
preliminarily  demonstrated in~\cite{lahaye05}, lies in the
production, by means of laser cooling techniques, of a
magnetically guided atomic beam on which evaporative cooling can
then be applied spatially. In this case the duty cycle of laser
cooling is typically 50\%. Another option consists in transferring
a cloud of cold atoms in a three-dimensional magnetic trap and
then move it away from the MOT, in order to load the latter again.
The trapped cloud can then be evaporated `on the fly' in this
moving trap, provided no heating and no spin-flip losses occur.

Magnetic transport has already been demonstrated in macroscopic
traps moving mechanically~\cite{tracks} or using a set of coils
with time-varying currents~\cite{Esslinger}. Those traps use a
three-dimensional quadrupole configuration with a vanishing field
at the center, which prevents evaporation to degeneracy. On atom
chips, the transport of ultracold~\cite{Reichel2001} and even of
Bose-condensed clouds~\cite{Reichel2005} has been demonstrated,
but still with limited fluxes.

\begin{figure}
\includegraphics[width=6cm]{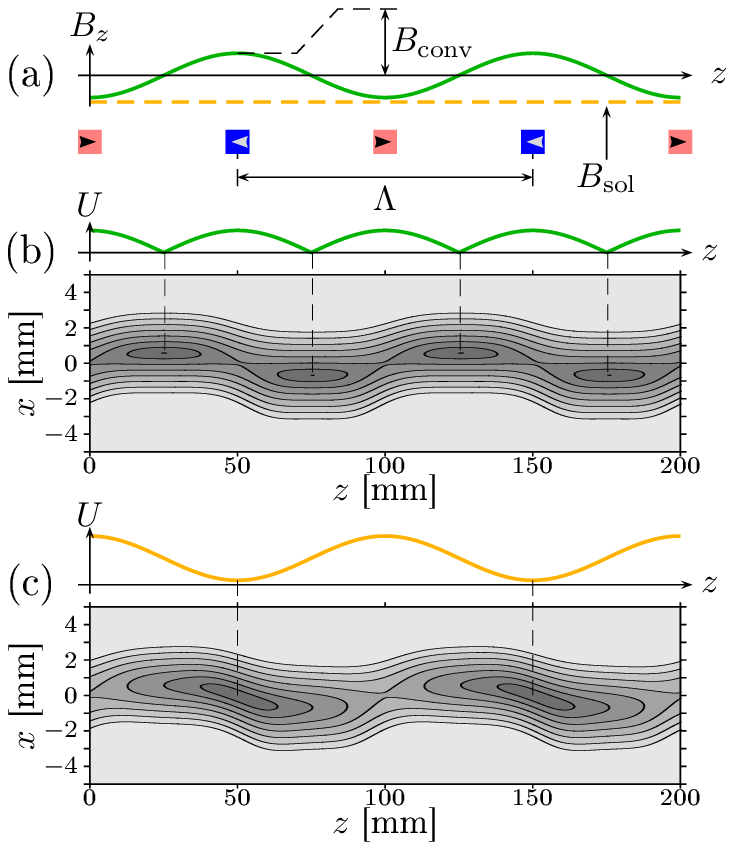}
\caption{(Color online). Creation of a longitudinal confinement
with permanent magnets. (a): Magnets with alternating
magnetization directions (arrows) located on a line parallel to
$z$ create a modulated $z$-component of the magnetic field (solid
line). A constant offset field $B_{\rm sol}$ (dashed line) can be
added in order to keep a constant sign for $B_z$. (b): The
potential $U$ experienced by the atoms for $B_{\rm sol}=0$ is a
chain of quadrupole traps (top curve). Due to the transverse
component $B_x$ of the field, the traps are shifted from the guide
axis, as can be seen on the contour plot of $|{\boldsymbol B}|$ in
the $(xz)$ plane. The iso-contours are plotted every 20~G,
starting at 0~G. (c): With a large enough $B_{\rm sol}$, one gets
a chain of on-axis Ioffe-Pritchard traps.} \label{fig:1}
\end{figure}

In this Letter, we demonstrate transport and evaporative cooling
of several atom clouds in a chain of moving Ioffe-Pritchard traps.
The transverse confinement is ensured by a magnetic guide creating
a two-dimensional quadrupole field, and a spatially varying
longitudinal field created by permanent magnets fixed on a
conveyor belt provides a moving longitudinal trapping. We show
that this setup can be used to transport an atom cloud at a speed
as low as 25~cm/s, without any detectable heating, and that we can
simultaneously transport and cool several clouds, each containing
about $10^9$ atoms.

The setup used to produce a magnetically guided atomic beam of
$^{87}$Rb in the collisional regime has been described
elsewhere~\cite{prlrb2}. Packets containing $2\times10^9$ atoms in
the $|F=-m_F=1\rangle$ state are injected every 200~ms into a
4.5~m long magnetic guide generating a transverse gradient
$b=800$~G/cm. Due to their longitudinal velocity dispersion, they
spread and overlap, resulting in a continuous beam after typically
50~cm of propagation.

A train of magnetic traps is obtained by superimposing a
corrugated bias field $B_z$ along the guide axis, which creates
longitudinal potential barriers. This is achieved with permanent
magnets located on a line parallel to the guide axis. In practice,
they are separated by a distance $\Lambda/2=5$~cm (see
Fig.~\ref{fig:1}a). The resulting potential consists in a chain of
three-dimensional quadrupole traps (Fig.~\ref{fig:1}b). By adding
with a solenoid a uniform longitudinal field $B_{\rm sol}$ higher
than the corrugation amplitude $B_{\rm conv}$, two successive
quadrupole traps merge into a single Ioffe-Pritchard (IP) trap
(Fig.~\ref{fig:1}c). For a given configuration, the IP trap depth
is twice the one of a quadrupole trap.

\begin{figure}
\includegraphics[width=8.5cm]{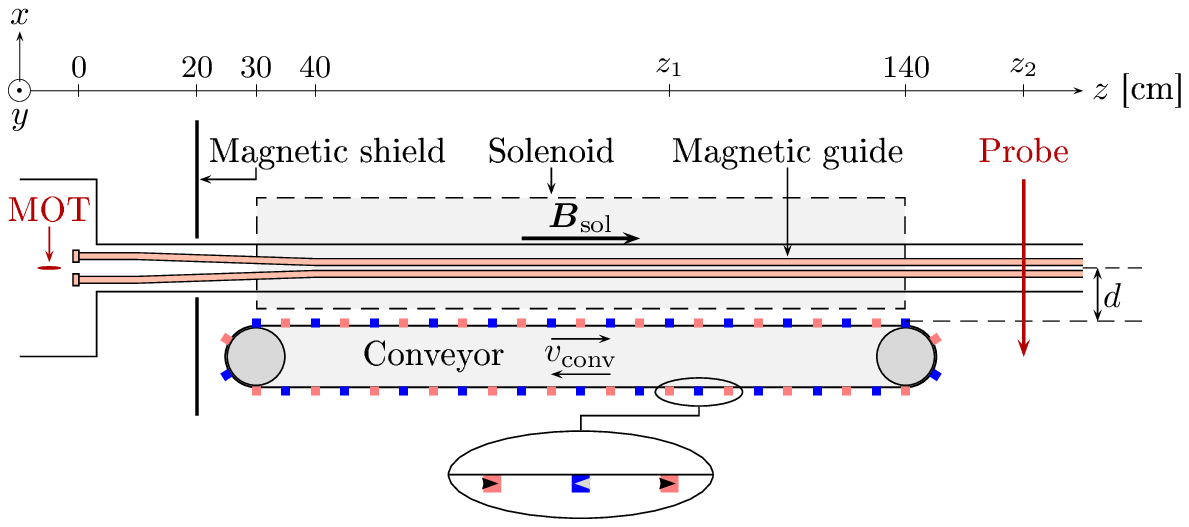}
\caption{(Color online). Experimental setup.} \label{fig:2}
\end{figure}

In order to let the resulting traps move along $z$ at a
controllable velocity $v_{\rm conv}$, the magnets are fixed on a
conveyor belt. The practical arrangement is shown schematically on
Fig.~\ref{fig:2}. The conveyor belt supporting 50~magnets is
parallel to the guide axis, at an adjustable distance $d$. This
allows us to vary the height of the magnetic field barriers
experienced by the atoms, since the corrugation amplitude $B_{\rm
conv}$ scales approximately as $\exp(-2\pi
d/\Lambda)$~\cite{hindsreview}. We use $20\times10\times10$~mm$^3$
rare-earth (Nd-Fe-B) permanent magnets~\cite{hulet} with a
magnetization of about 800~kA/m, yielding a field amplitude
$B_{\rm conv}\simeq25$~G for a distance $d\simeq45$~mm. The
resulting IP traps have a depth of $2B_{\rm conv}=50$~G, a typical
offset field $B_{\rm sol}-B_{\rm conv}\simeq1$~G, a transverse
gradient $b\simeq800$~G/cm, and a longitudinal curvature on the
order of 10~G/cm$^2$. The corresponding axial and transverse
frequencies are 3 and 720~Hz, respectively. The barrier height
increases fast at the entrance of the conveyor (for $z<30$~cm),
which permits an efficient capture of atom clouds launched into
the guide.

The conveyor is set in motion by an electric motor and its speed
$v_{\rm conv}$ can be adjusted between 10 and 130~cm/s. The
measured speed fluctuations are below 2\%. The depths of
successive wells vary by as much as 5\% due to the dispersion in
the magnetization of the magnets. However, by contrast to magnetic
transport systems using time-varying currents in a set of coils,
the geometry of a given trap, in the moving frame, is constant by
construction, even far from the trap minimum. This prevents any
strong heating due to the deformations of the trap during the
motion. The MOT region is protected by a magnetic shield from the
influence of the conveyor and solenoid fields (see
Fig.~\ref{fig:2}).

\begin{figure}
\includegraphics{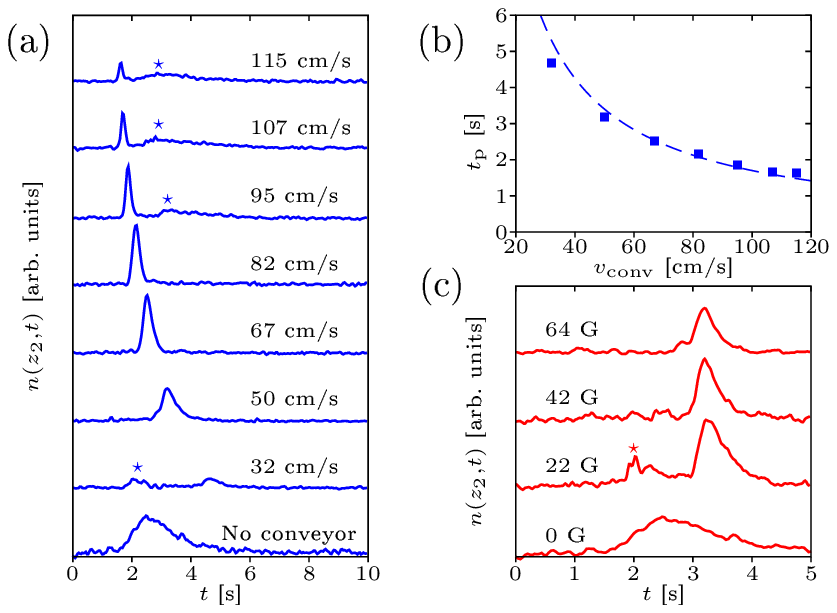}
\caption{(Color online). Absorption of a probe beam located at
$z_{2}=1.7$~m from the guide entrance as a function of time $t$
elapsed after the launch of a single atom packet, at the velocity
$v_{\rm i}=80$~cm/s (with a rms velocity dispersion of 20~cm/s).
(a): The IP traps have a depth of 50~G and a minimum field of 1~G,
and one varies $v_{\rm conv}$. A sharp peak of trapped atoms
arrives at a time $t_{\rm p}$. When the velocity mismatch between
the conveyor and the injected atoms is too large, extra peaks
corresponding to untrapped atoms appear~($\star$). (b): Arrival
time $t_{\rm p}$ of the trapped atoms (squares) as a function of
$v_{\rm conv}$; $t_{\rm p}$ is essentially equal to $z_{2}/v_{\rm
conv}$ (dashed line). (c): The conveyor speed is fixed at 50~cm/s
and one varies the IP trap depth. For too shallow traps, one gets
a peak of untrapped atoms ($\star$); if the depth is too high,
some atoms are reflected, which decreases the peak amplitude.}
\label{fig:3}
\end{figure}

We first investigate the transport of a single atom cloud launched
into the potential resulting from the guide and the conveyor. With
a resonant probe located at $z_{2}=1.7$~m from the guide entrance,
we mesure the time-dependant atomic density $n(z_{2},t)$.
Figure~\ref{fig:3}a shows the corresponding signals for a chain of
IP traps with a 50~G depth and a minimum of 1~G. The velocity of
the packet before entering the conveyor is $v_{\rm i}=80$~cm/s,
with a rms dispersion of $20$~cm/s, and we study different
conveyor velocities $v_{\rm conv}$. Without conveyor (bottom
curve), the temporal width of the signal exceeds one second, due
to the large spreading of the packet during its free flight. When
the conveyor is running at a velocity around $v_{\rm i}$, one gets
a sharp peak arriving at the time $t_{\rm p}\sim z_{2}/v_{\rm
conv}$ (Fig.~\ref{fig:3}b) corresponding to atoms that have been
trapped in the conveyor. Indeed, for those atoms, the spreading is
frozen out during the transport, thus avoiding a decrease in the
atomic density. The width of the absorption signal is set by the
residual spreading over 30~cm once the atoms are released from the
conveyor.

A significant fraction of the atoms are captured even if $v_{\rm
conv}$ differs significantly from $v_{\rm i}$ (see, \emph{e.g.},
the curve for $v_{\rm conv}=50$~cm/s). For a large velocity
mismatch, one observes a class of atoms that are not trapped. If
$v_{\rm i}>v_{\rm conv}$, these untrapped atoms are too energetic
to be trapped, `fly' over the longitudinal barriers, and arrive at
the probe location before the trapped ones. Conversely, if $v_{\rm
i}<v_{\rm conv}$, one observes a peak arriving at large times,
corresponding to slow atoms. The best capture efficiency is
obtained for $v_{\rm conv}\simeq v_{\rm i}$. The lowest transport
velocity we have been able to achieve is as low as $v_{\rm
conv}=25$~cm/s, for which collisions with the background gas start
to decrease significantly the number of trapped atoms reaching the
probe region.

\begin{figure}
\includegraphics[width=8cm]{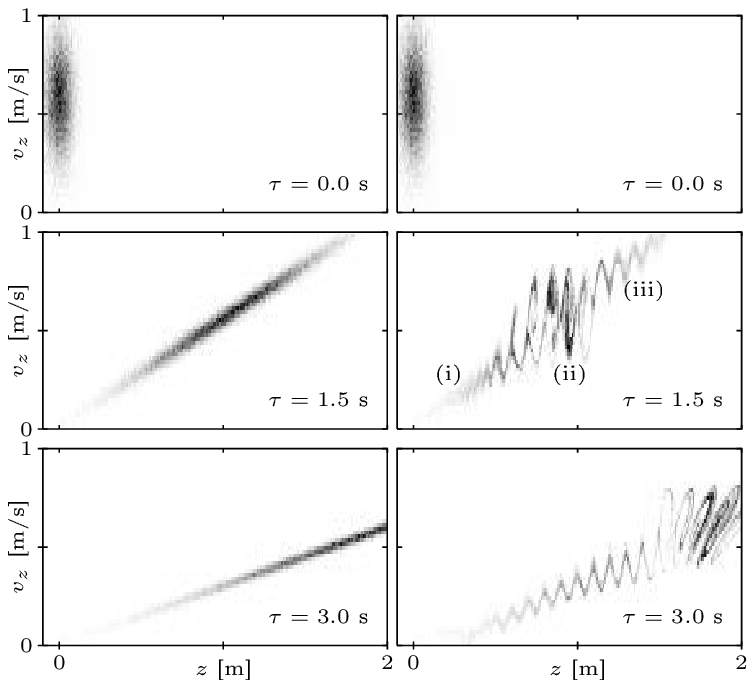}
\caption{Simulated atomic distribution in phase-space $(z,v_z)$
for various times $\tau$ after the launch of the packet. The
initial velocity is $v_{\rm i}=60$~cm/s, with a 20~cm/s
dispersion, and the initial size of the packet is 5~cm. Left: free
propagation (no conveyor), showing the spatial spreading of the
packet and the building of correlations between position and
velocity. Right: the conveyor is made of IP traps with 10~G
height, moving at $v_{\rm conv}=60$~cm/s. At $\tau=1.5$~s, the
slow (i), trapped (ii) and fast (iii) atoms are easily identified.
At $\tau=3$~s, the atom packets released from the conveyor have
already started to spread out and overlap.} \label{fig:4}
\end{figure}

We then study the influence of the depth of the conveyor potential
on the capture. Figure~\ref{fig:3}c depicts the signal obtained
when launching atoms at $v_{\rm i}=80$~cm/s into a conveyor
running at $v_{\rm conv}=50$~cm/s, for various depth of the IP
wells: 22, 42 and 64~G. If the barrier heights are too low, one
clearly distinguishes a peak corresponding to fast, untrapped
atoms. When raising the barrier heights, this peak disappears. If
the barriers are too high, the overall signal starts to decrease
as many atoms are reflected at the conveyor entrance.

For a better understanding of the trapping process, we have
developed a numerical simulation of the atomic trajectories in the
conveyor belt potential. It gives results in very good agreement
with the observed arrival time signals and confirm that, depending
on the barrier height and on the velocity mismatch between the
conveyor and the injected packets, atoms can: (i) be considerably
slowed down during the entrance (low $v_{\rm i}$, large height),
some of them being even reflected; (ii) be trapped in the conveyor
wells; and (iii) be too energetic to be trapped and simply pass
over the conveyor barriers (large $v_{\rm i}$, low height).
Figure~\ref{fig:4} presents plots of the atomic distribution in
the phase-space $(z,v_z)$, at a time $\tau$ after the launch of
the packet, obtained by numerical simulation. These plots allow
for an easy identification of the various classes of atoms. The
slowing of the atoms of class (i) by the time-dependent potential
is reminiscent of the Stark deceleration technique used for beams
of polar molecules~\cite{meijer}, and can be understood
qualitatively as the result of a reflection of the atoms on a
moving potential wall~\cite{neutron}. It is clear from such plots
that the spatial spreading of the trapped clouds is frozen during
the transport. Here, the initial size and velocity spread of the
packet are such that several conveyor wells are loaded.

\begin{figure}
\includegraphics[width=6cm]{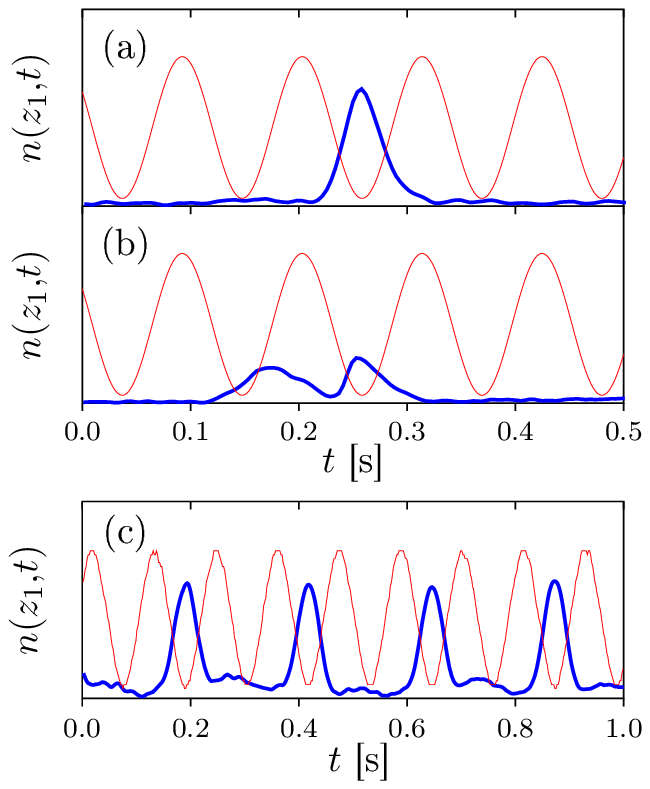}
\caption{(Color online). Absorption signal (thick line) measured
\emph{in the conveyor belt}, with a probe located at $z_{1}=1$~m.
The IP traps have a depth of 32~G, and one has $v_{\rm
conv}=v_{\rm i}=88$~cm/s. The thin line is the measured $B_z$
field component produced by the conveyor. (a): Proper
synchronization of the launching of one packet to load a single
trap (the cloud arrives in coincidence with a potential minimum).
(b): The launching occurs at a time $\Lambda/(2v_{\rm conv})$
before, resulting in a splitting of the cloud into two consecutive
traps. (c): Multiple injection of packets, with the proper
synchronization to load one every second trap.} \label{fig:5}
\end{figure}

For an optimal loading of a single trap, three conditions need to
be fulfilled. First, our simulations confirm that, as one expects
intuitively, the energy of the trapped atoms in the conveyor frame
is minimum when the velocities are matched $v_{\rm i}\simeq v_{\rm
conv}$. Second, the length of the packet at the entrance of the
conveyor has to be smaller than the distance between adjacent
traps. Finally, a careful synchronization of the launching with
respect to the conveyor motion is required. This is essential to
avoid a splitting of the cloud between two adjacent wells (see
Fig.~\ref{fig:5}a and b). When those conditions are fulfilled,
typically $N\sim10^9$ atoms are trapped, which corresponds,
according to our simulations, to 75\% of the incoming packet.

We now turn to the injection of multiple packets into the conveyor
belt. In order to demonstrate the trapping efficiency of the
conveyor, we use the probe at $z_{1}=1$~m (\emph{i.e.}, inside the
conveyor zone) and inject atoms with the proper synchronization in
order to capture one cloud every second IP trap
(Fig.~\ref{fig:5}c). When released from the conveyor, those
packets spread out and overlap, yielding a continuous beam. We
investigated the effect of the conveyor on the beam's temperature,
and find no detectable heating within our experimental accuracy:
for $v_{\rm conv}=v_{\rm i}=1$~m/s, one has $T=600\pm20$~$\mu$K
without conveyor, and $590\pm20$~$\mu$K with a conveyor having a
35~G height. This result is compatible with our numerical
simulations: for our parameters, the heating ($\sim 10$~$\mu$K)
associated with the presence of the conveyor is negligible with
respect to the initial temperature. The proper synchronization of
the injection of packets is limited in our current setup to
velocities above $\sim80$~cm/s. For lower velocities, the
longitudinal size of the packet at the conveyor entrance already
exceeds the distance $\Lambda$ between adjacent IP traps, due to
the non-adiabaticity of the entrance into the guide, and to the
transverse compression of the confining potential over the first
40~cm of the guide (see Fig.~\ref{fig:2}).

This chain of IP traps constitutes a very simple way of
transporting and cooling in parallel many atomic clouds. The next
step is to combine this transport with evaporative cooling, thus
paving the way for the parallel production of Bose-Einstein
condensates as well as for the achievement of a cw atom laser. For
our parameters, the collision rate within one trapped packet is
already on the order of $10~{\rm s}^{-1}$, which has allowed for a
first demonstration of evaporation. Using radio-frequency fields,
we have been able to remove selectively the untrapped atoms and,
moreover, to decrease the beam's temperature by a factor of two,
reaching $280\pm10$~$\mu$K, with a flux reduction by a factor of
four. Those preliminary results are encouraging in view of the
realization of a new experimental set-up, designed on purpose for
the use of such a scheme.

In view of the achievement of a cw atom laser through direct
evaporation of an atomic beam~\cite{Mandonnet00}, the use of a
train of IP traps combined with evaporation would allow for the
realization of an ultra-slow, but still supersonic, atomic beam
(the latter condition being essential in order not to decrease
drastically the atomic flux~\cite{vogels}). For that purpose, one
would capture packets of atoms at low speed into the conveyor, and
then compress them adiabatically by increasing the strength of the
transverse confinement. The resulting hot and dense clouds can
then be evaporatively cooled. An interesting strategy consists in
reaching a temperature $T$ that satisfies $k_{\rm B}T\ll mv_{\rm
conv}^2$, so that the packets can be released into the guide to
overlap and form a very slow continuous supersonic beam. Compared
to the direct injection of packets into the guide, (i) the effect
of the initial longitudinal dilution due to the spreading of
packets is minimized, (ii) the efficiency of the 3D evaporation is
higher than its 2D counterpart~\cite{ketterlereviewevap}, and
(iii) the time available for evaporation is increased
considerably, as the beam's velocity is set by $v_{\rm conv}$,
which can be as low as $10$~cm/s. The corresponding final
temperature then needs to be smaller than 10~$\mu$K. One could
apply, on the atomic beam obtained this way, the final evaporation
stages in order to achieve quantum degeneracy.

We thank Jean Dalibard for a careful reading of the manuscript,
and the ENS laser cooling group for fruitful discussions. We
acknowledge financial support from the D\'el\'egation G\'en\'erale
pour l'Armement (DGA) and Institut Francilien de Recherche sur les
Atomes Froids (IFRAF). Z.~W. acknowledges support from the
European Marie Curie Grant MIF1-CT-2004-509423, and G.~R. support
from the DGA.

\end{document}